\documentclass[journal]{IEEEtran}
\usepackage{mathrsfs}
\usepackage{amssymb}
\usepackage{bbm}
\usepackage{amsfonts}
\usepackage{cite}
\usepackage{comment}
\usepackage{setspace}
\usepackage{graphicx}
\usepackage{psfrag}
\usepackage{subfig}
\usepackage{amsmath}
\usepackage{bm}
\usepackage{algorithm}
\usepackage{algorithmic, algorithmic-fix}

\DeclareMathOperator*{\argmax}{arg\,max}

\begin{document}
\title{Enhanced Uplink Resource Allocation in Non-Orthogonal Multiple Access Systems}

%

%

\author{Rukhsana Ruby, Member, IEEE, Shuxin Zhong, Hailiang Yang, and Kaishun Wu, Member, IEEE

\thanks{This research was supported in part by the China NSFC Grant 61472259, Guangdong Natural Science Foundation (No. 2017A030312008, 2016A030313036), Shenzhen Science and Technology Foundation (No. JCYJ20170302140946299, JCYJ20170412110753954, JCYJ20150324140036842).Guangdong Talent Project 2014TQ01X238, 2015TX01X111, GDUPS(2015). Kaishun Wu is the corresponding author. R. Ruby, S. Zhong, H. Yang and K. Wu are with the college of Computer Science and Software Engineering, Shenzhen University, Shenzhen, Guangdong, 518060 China, e-mail: ruby@szu.edu.cn
}
}

\maketitle

\begin{abstract}

Non-orthogonal multiple access (NOMA) is envisioned to be one of the most beneficial technologies for next generation wireless networks due to its enhanced performance compared to other conventional radio access techniques. Although the principle of NOMA allows multiple users to use the same frequency resource, due to decoding complication, information of users in practical systems cannot be decoded successfully if many of them use the same channel. Consequently, assigned spectrum of a system needs to be split into multiple subchannels in order to multiplex that among many users. Uplink resource allocation for such systems is more complicated compared to the downlink ones due to the individual users' power constraints and discrete nature of subchannel assignment. In this paper, we propose an uplink subchannel and power allocation scheme for such systems. Due to the NP-hard and non-convex nature of the problem, the complete solution, that optimizes both subchannel assignment and power allocation jointly, is intractable. Consequently, we solve the problem in two steps. First, based on the assumption that the maximal power level of a user is subdivided equally among its allocated subchannels, we apply many-to-many matching model to solve the subchannel-user mapping problem. Then, in order to enhance the performance of the system further, we apply iterative water-filling and geometric programming two power allocation techniques to allocate power in each allocated subchannel-user slot optimally. Extensive simulation has been conducted to verify the effectiveness of the proposed scheme. The results demonstrate that the proposed scheme always outperforms all existing works in this context under all possible scenarios.

\end{abstract}

\IEEEpeerreviewmaketitle

\section{Introduction}


\IEEEPARstart{T}he data traffic over cellular networks is projected to grow explosively in the coming years due to the proliferation of smartphones, tablets, smart terminals and emerging applications (e.g., machine-type-communications (MTC))~\cite{STekinay2002, Borst2005, ArunAgarwal2014, AAli2015}. Consequently, future radio access networks~\cite{SGhosh2005, VSuryaprakash2016} are expected to have the capability of supporting massive connectivity, diverse sets of users and applications with radically different requirements in terms of delay, bandwidth and so on. In order to obtain fruitful outcome in this context, designing an effective and efficient radio access technology~\cite{TEdler2014} is one of the possible solutions. Through experimentation and theoretical analysis, it is proved that non-orthogonal multiple access (NOMA) technique is able to provide enhanced performance comparing with other orthogonal multiple access (OMA) techniques, such as time division multiple access (TDMA) and frequency division multiple access (FDMA)~\cite{Higuchi2013, YEndo2012, JUmehara2012, NOtao2012}. Consequently, NOMA is considered as the future dominating radio access technique, and is expected to satisfy the ever-increasing demands of future cellular networks. 

Conceptually, power-domain NOMA allows multiple users to occupy the same frequency channel. By applying successive interference cancellation (SIC)~\cite{NIMiridakis2013} in NOMA systems, superposition coded signal can be correctly decoded and demodulated at the receiver. Although NOMA technique allows multiple users to be superimposed on the same frequency channel, due to the error propagation in the SIC technique, it is not an optimal design to assign large number of users on the same channel. Consequently, dedicated spectrum of a system needs to be subdivided into multiple subchannels in order to support increased number of users. At the same time, how to allocate these subchannels among users in a multiplex manner given the allowable maximum number of users that can use a subchannel simultaneously, is an important problem. Extensive research has been conducted on the downlink subchannel and power allocation for such NOMA systems. Based on some assumption of having constant power on the subchannels, typically, existing works provide some heuristics for subchannel-user mapping task. Once the subchannel-user mapping information is known, in order to enhance the performance of the system further, different existing works have provided different schemes for power allocation across the allocated subchannel-user slots. For example, in \cite{YSaito2013, ABenjebbour2013}, the authors use fractional transmit power allocation technique among users and equal power allocation concept across subchannels.~\cite{MRHojeij2015} uses water-filling-based approach for power allocation, and in \cite{PParida2014}, the authors use difference of convex (DC) programming-based~\cite{NVucic2010} approach for the power allocation in both user and subchannel levels. Energy-efficient downlink resource allocation has also been studied in some papers, such as \cite{SHan2014, QSun2015, FFang2016}.   

Unlike the downlink one, uplink resource allocation even in conventional OMA systems is considered challenging~\cite{JHuang2009, RRuby2014, RRuby2015} because of the individual users' power constraints and discrete nature of subchannel assignment. On the other hand, decoding technique in NOMA systems, SIC, is a multi-user detection technique that uses the structured nature of interference to decode multiple concurrent transmissions. Each individual signal from the composite signal is retrieved one by one following some order. If any of the signals is failed to be decoded, it is unlikely that the rest of the signals can be decoded. Therefore, decoding order plays the significant role on the success of decoding operation and the throughput achieved by each individual signal. All these complications bring further difficulties in the uplink resource allocation of NOMA systems compared to OMA systems. Although NOMA principle does not enforce the decoding order of received superposition coded signals, it is proved in~\cite{MMollanoori2014} that decoding of stronger signals ahead of weaker signals is beneficial for the system in terms of throughput and proportional fairness.

Compared to the downlink resource allocation in NOMA systems, uplink resource allocation is not that much studied. Still, there are some works in this context. Unlike the system in our work (in which NOMA technique is employed in the frequency domain), for a system in which NOMA technique is employed in the time domain, a set of uplink resource allocation schemes is provided in~\cite{KKumaran2003, MMollanoori2014} with the objective of throughput maximization and fairness of the system. Although the scheduling scheme in~\cite{KKumaran2003} assumes that the system has one time slot and a set of users with their power constraints to schedule, the scheduling scheme in~\cite{MMollanoori2014} optimizes the total throughput and fairness of the system over a set of time slots and users. On the other hand, the main drawback of the work in~\cite{MMollanoori2014} is, the time slots are resource elements and are invariant over time, which is very impractical for wireless networks. Moreover, given the power constraint of each individual user, each user can get only one time slot (i.e., one resource block), the concept of which fails to exploit multi-user-channel diversity of wireless systems. However, in practice, if multiple time slots are allocated to a user, the performance of that user may be enhanced. Another very close work compared to our work is~\cite{MAlImari2015}. In this work, the authors have proposed an uplink subchannel and power allocation scheme based on iterative water-filling technique~\cite{WeiYu2004}. With the expectation of utilizing the multi-user-channel diversity, this resource allocation scheme overcomes the drawback of the solution in~\cite{MMollanoori2014} by assigning multiple subchannels (i.e., multiple resource blocks) to each user. Moreover, this scheme assigns exactly the maximum allowable number of users to each subchannel and gives more preference to the users with better channel while solving the subchannel-user mapping problem. However, even in uplink OMA systems~\cite{RRuby2014}, we previously observed that not necessarily the more the users allocated to each subchannel, the better the throughput is, especially in worse channel condition. This is because each user requires to subdivide its limited power level among its allocated subchannels. Furthermore, in NOMA systems, giving less privilege to the users with worse channel not necessarily enhances the throughput. Since the power level of other users in such systems is considered as interference for some particular user, pairing users with highly different channel condition is sometimes conducive to the performance of the system. In this paper, our objective is to overcome the drawbacks of existing aforementioned uplink resource allocation schemes, and to take NOMA-specific all useful scheduling insights into account.   

The contribution of this paper is an elegant uplink subchannel and power allocation scheme in a NOMA system with enhanced performance. Since this problem considers subchannels assignment which are associated with discrete variables in the formulated problem, the problem is NP-hard. Moreover, even if the subchannel assignment information is known, because of the interference power resultant from the superposition coded signals of other users on a specific subchannel, the power allocation of the problem is non-convex~\cite{DPBertsekas1999}. As a result, joint subchannel assignment and power allocation of this problem can be considered as a mixed integer non-linear programming (MINLP) problem. Overall, joint optimization of both subchannel assignment and power allocation is not tractable for this case. Consequently, we solve this problem in two steps. Based on the assumption that the maximal power level of each user is subdivided equally among its allocated users, we apply many-to-many matching model~\cite{KHamidouche2014, ARoth1984} to solve the subchannel-user mapping problem. Then, we apply iterative water-filling~\cite{WeiYu2004} and geometric programming (GP)~\cite{MChiang2005} techniques to allocate power across the subchannels for different users. Iterative water-filling is a multi-user-channel power allocation technique, which is developed based on the insights of single-user water-filling solution. On the other hand, GP technique can solve special-form of non-convex problems using convex optimization solvers through variable transformation. Given the subchannel-user mapping information, our uplink power allocation problem is amenable to GP after applying some transformation on the objective function. Extensive simulation has been conducted to verify the effectiveness of our proposed uplink resource allocation scheme comparing with two very similar existing works~\cite{MMollanoori2014, MAlImari2015}. The results demonstrate that the proposed scheme always outperforms the existing works in terms of computational complexity, the usage of resource and overall performance.  


The rest of the paper is organized as follows. Along with the background information and the description of the system, we formulate our uplink resource allocation problem in Section~\ref{sec:sysmodel}. The detailed solution approach is provided in Section~\ref{sec:sol}. Followed by the simulation methodology, we evaluate the performance of our uplink resource allocation scheme in Section~\ref{sec:eval}. Finally, Section~\ref{sec:concl} concludes the paper with some direction on future research.

\section{System Model and Problem Formulation}
\label{sec:sysmodel}

In this paper, we consider an uplink scenario of a cellular network, which has one base station. Time is divided into frames, and in each frame, the entire pre-assigned spectrum for the system is divided into $N$ subchannels with equal bandwidth. The resultant subchannels are the elements of a set, denoted by \textbf{N}. There are $M$ number of users in the system, and the corresponding set holding these users is denoted by \textbf{M}. Using the subchannels in set \textbf{N} as the transmission media, the users in set \textbf{M} transmit data to the base station. Each user $m$ in set \textbf{M} has the maximal power level, denoted by $p_m^{\mbox{max}}$. Both the base station and the users in the system are equipped with NOMA technologies. The users transmit their data using superposition coding (SC) technique over a set of subchannels. Whereas, the receiver, i.e., the base station applies SIC technique on each subchannel to decode the superimposed signals, and extracts the corresponding signal of each individual user. However, before the uplink transmission operation, it is required to schedule subchannels and power across the users optimally so that the capacity of the system is maximized. We assume that the scheduling scheme in the system is centralized, and the base station is appointed to conduct this operation. To develop this scheduling scheme, the entire channel state information (CSI) of the system is required, and hence the base station is aware of all these information. At the beginning of each time frame, all users send their CSI to the base station via some reliable control channels. 

We consider the block fading channel model~\cite{DavidTse2005}. It implies that the CSI of the subchannels in the system remains constant over a time frame, however vary independently across different time frames. Although NOMA techniques have various classification, we plan to exploit power-domain NOMA. We assume that the base station assigns $M_n$ number of users on the $n$th subchannel, and the corresponding set holding these users is denoted by $\textbf{M}_n$. If each individual user $m$ transmits $\sqrt{p_m^n}{s_m}$ symbol on subchannel $n$, the symbol received by the base station on this subchannel can be expressed as

\begin{equation}
    x_n = \displaystyle\sum_{m=1}^{M_n}\sqrt{p_m^n}{s_m},
\end{equation}

\noindent
where $s_m$ is the modulated symbol of the $m$th user on subchannel $n$, and $p_m^n$ is the power level assigned to user $m$ on subchannel $n$. Consequently, the signal of user $m$, received by the base station on subchannel $n$, can be represented as

\begin{eqnarray}
\nonumber & y_m^n = h_m^nx_n + z_n \\
& = \sqrt{p_m^n}h_m^ns_m + \displaystyle\sum_{i=1, i \neq m}\sqrt{p_i^n}h_m^ns_i + z_n,
\end{eqnarray}

\noindent 
where $h_m^n$ is the channel gain of user $m$ on the $n$th subchannel. $z_n$ is the noise power over subchannel $n$, which follows Additive White Gaussian Noise (AWGN)~\cite{KMcClaning2000} distribution with mean zero and variance ${\sigma}_n^2$, i.e., $z_n \approx {\cal{CN}}(0, {\sigma}_n^2)$. The noise power of subchannel $n$ is statistically same for all users. In NOMA systems, each subchannel is shared by multiple users. Consequently, each user on subchannel $n$ receives its signal as well as the interference signals from other users on the same subchannel. Therefore, without SIC at the base station, the received SINR of the $m$th user on subchannel $n$ is given by

\begin{equation}
\mbox{SINR}_m^n = \frac{p_m^n|h_m^n|^2}{{\sigma}_n^2 + \displaystyle\sum_{i=1, i \neq m}^{M_n}p_i^n|h_i^n|^2} = \frac{p_m^n{g_m^n}}{1 + \displaystyle\sum_{i=1, i \neq m}^{M_n}p_i^ng_i^n},
\end{equation}

\noindent
where ${\sigma}_n^2 = E[|z_n|^2]$ is the noise power on subchannel $n$, and $g_i^n = |h_i^n|^2/{{\sigma}_n^2}$ is the normalized channel gain of user $m$ on subchannel $n$. Based on Shannon's capacity formula~\cite{HMichiel2001}, the sum-rate of subchannel $n$ is given by

\begin{eqnarray}
\label{eq:sum-rate-wo-sic}
\nonumber & R_n = \displaystyle\sum_{m=1}^{M_n} \mbox{log}_2\left(1 + \mbox{SINR}_m^n\right) \\
& = \displaystyle\sum_{m=1}^{M_n} \mbox{log}_2\left(1 + \frac{p_m^n{g_m^n}}{1 + \displaystyle\sum_{i=1, i \neq m}^{M_n}p_i^ng_i^n}\right).
\end{eqnarray}

\begin{figure}
  \begin{center}
    \includegraphics[width=0.6\columnwidth]{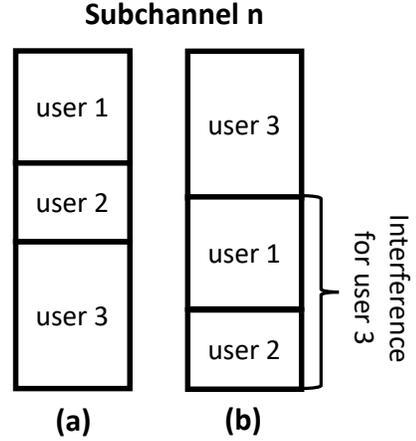}
    \caption{A sample example for the organization of decoding order in subchannel $n$ ((b) is the organized version of (a)), where $p_1^ng_1^n = 20, p_2^ng_2^n = 10~\mbox{and}~p_3^ng_3^n = 30$ ($1$st case), or $g_1^n = 0.2, g_2^n = 0.1~\mbox{and}~g_3^n = 0.3$ ($2$nd case).}
    \label{fig:decoding-order}
  \end{center}
\end{figure}

In NOMA systems, the SIC process is implemented at the receiver to reduce the interference from other users on the same subchannel. According to~\cite{MMollanoori2014}, it has been proved that the optimal decoding order is equivalent to the decreasing order of received power. In this way, the interference imposed on each user by other users of the same subchannel is reduced, and consequently the sum-capacity and proportional fairness of the system are enhanced. Based on this truth, we adopt the following rule while determining the decoding order of the users transmitting on the same subchannel. As we mentioned in the introduction that our solution approach of this problem consists of two steps. In the first step, based on the assumption that the maximal power level of each user is equally subdivided among its allocated subchannels. In this case, we decode the users superimposed on a subchannel in the decreasing order of their received power. A sample example of this idea has been provided in Fig.~\ref{fig:decoding-order}.

For the second case, when the user and subchannel assignment information are known, the power allocation of the superimposed users should follow some order. From the insights of the decoding order idea of~\cite{MMollanoori2014} as well as intuitively, it is obvious that the user with better gain in any subchannel should be assigned with larger transmit power. In this way, the interference imposed on any user (assigned to any subchannel) caused by other users is reduced. A sample example of this idea is provided in Fig.~\ref{fig:decoding-order} as well. Due to the decoding order concept and the principle of SIC technique, not necessarily other all users assigned to a particular subchannel impose interference on a specific user. On subchannel $n$, denote that the users in set $\textbf{M}_n^m$ produce interference for user $m$. Consequently, the sum-rate of subchannel $n$ (in (\ref{eq:sum-rate-wo-sic})) can be rewritten as

\begin{equation}
\label{eq:Rn}
R_n = \displaystyle\sum_{m=1}^{M_n} \mbox{log}_2\left(1 + \frac{p_m^n{g_m^n}}{1 + \sum_{i \in \textbf{M}_n^m}p_i^ng_i^n}\right).
\end{equation}

In this work, while preserving the power constraints of all users, our objective is to allocate the subchannels in set \textbf{N} across all users in set \textbf{M}, and assign power level to each subchannel-user slot so that the capacity of the system is maximized. Clearly, this is an optimization problem. To formulate this problem, we define a binary variable ${\alpha}_m^n$. ${\alpha}_m^n = 1$ implies that subchannel $n$ is allocated to user $m$, and ${\alpha}_m^n = 0$ means the other case.  It is proved in~\cite{MAlImari2015} that the more users are assigned to a subchannel, the better the system capacity is. However, due to the varying nature of wireless channels and the decoding complication of SIC technique, not necessarily more users assigned to a subchannel will enhance the system throughput. While giving weight to this observation and insight, we assume that maximum $K$ number of users can be assigned to a subchannel. Therefore, the uplink subchannel and power allocation problem in this context can be formulated as follows.

\begin{eqnarray}
\label{eq:opt-prob1} &\displaystyle\max_{\{p_m^n, {\alpha}_m^n\}}\displaystyle\sum_{n \in \textbf{N}}\displaystyle\sum_{m \in \textbf{M}_n}{\alpha}_m^n\mbox{log}\left(1 + \frac{p_m^ng_m^n}{1 + \displaystyle\sum_{i \in \textbf{M}_n^m}p_i^ng_i^n}\right), \\
&\nonumber \mbox{subject to,} \\
\label{eq:opt-prob2} & \displaystyle\sum_{m \in \textbf{M}_n}{\alpha}_m^n \le K,~ \forall{n \in \textbf{N}}, \\
\label{eq:opt-prob3} & \displaystyle\sum_{n \in \textbf{N}}{\alpha}_m^n \le \infty,~ \forall{m \in \textbf{M}}, \\
\label{eq:opt-prob4} & {\alpha}_m^n \in \{0, 1\},~ \forall{m \in \textbf{M}_n},~ \forall{n \in \textbf{N}}, \\
\label{eq:opt-prob5} & \displaystyle\sum_{n \in \textbf{N}}{\alpha}_m^np_m^n \le p_m^{\mbox{max}},~ p_m^n \ge 0,~ \forall{m \in \textbf{M}}.
\end{eqnarray}

In the above formulation, there are two types of variables, i.e., $\{{\alpha}_m^n\}$ and $\{p_m^n\}$. $\{{\alpha}_m^n\}$ are the set of discrete variables, and the problem is NP-hard because of these variables. On the other hand, even if the information of set $\{{\alpha}_m^n\}$ is known, because of the interference term $1 + \sum_{i \in \textbf{M}_n^m}p_i^ng_i^n$ inside the log term of (\ref{eq:opt-prob1}), the problem is non-convex. While considering the overall structure, we can say that the problem is jointly NP-hard and non-convex.


\section{Solution Approach}
\label{sec:sol}

In this section, we explore the solution approach of the uplink resource allocation problem of a NOMA system described in the previous section. The entire problem is formulated in~(\ref{eq:opt-prob1})-(\ref{eq:opt-prob5}). Apparently, due to the discrete nature of subchannel assignment (i.e., variables ${\alpha}_m^n$) and the continuous nature of power assignment (i.e., variables $p_m^n$), this is a MINLP problem. This type of problem even in conventional OMA systems is intractable. When it comes to the case of NOMA systems, due to the superimposition of multiple users on the same subchannel, the solution of this problem brings further complication. Therefore, we have decomposed the problem in two parts. In the first part, based on the assumption that the maximal power level of a user is subdivided equally among its allocated subchannels, we solve the subchannel-user mapping problem. In this case, we find that two-sided matching model is appropriate to capture the structure of this problem. Since one user can be assigned with multiple subchannels and one subchannel can have multiple users, many-to-many matching scheme is expected to solve the first problem. For the solution of the second part of the problem, we assume that we have the subchannel-user mapping information. Even though the subchannel-user mapping information is known, the power allocation across the subchannels and users, given the power constraints of the users, is a non-convex problem. Based on the structure of the problem, we find that iterative water-filling and GP are very appropriate techniques to solve this problem.

\subsection{Subchannel and User Mapping Scheme} 

In this system, intuitively, assignment of many users to a subchannel and allocating multiple users to a subchannel (to follow the guidelines of NOMA technique) is envisioned to enhance the overall throughput. Given the power constraint of each user, this problem is NP-hard. However, the nature of the problem implies that many-to-many matching model~\cite{KHamidouche2014, ARoth1984} is appropriate to solve this problem. Given that maximum $K$ number of users can be multiplexed on a subchannel, $M$ users in set \textbf{M} and $N$ subchannels in set \textbf{N} are two sets of players of this many-to-many matching relation. Note that each user $m$ can have infinite ($N$ in practice) number of users if possible. However, since user $m$ has maximal power constraint $p_m^{\mbox{max}}$, this should be subdivided equally among its allocated subchannels. 

\noindent
\textbf{Definition 1:} A many-to-many matching $\mu$ is a mapping from set \textbf{M} to set \textbf{N} such that every $m \in \textbf{M}$ and $n \in \textbf{N}$ satisfy the following properties:

\begin{itemize}

\item ${\mu}(m) \subseteq \textbf{N}$ and ${\mu}(n) \subseteq \textbf{M}$

\item $|{\mu}(m)| \le \infty, \forall{m \in \textbf{M}}$

\item $|{\mu}(n)| \le K, \forall{n \in \textbf{N}}$

\item $n \in {\mu}(m)$ if and only if $m \in {\mu}(n)$

\end{itemize}

\noindent
where ${\mu}(m)$ is the set of partners for user $m$ and ${\mu}(n)$ is the set of partners for subchannel $n$ under the matching model $\mu$. The definition states that each user in set \textbf{M} is matched to a subset of subchannels in set \textbf{N}, and vice versa. In other words, each user may choose a set of subchannels as the communication media, whereas each subchannel may choose a set of users to be assigned with in order to maximize the overall benefit of the system. However, before accomplishing these assignment operations, each user needs to have preference list based on some criteria. The criterion of constructing preference list for users is based on their received power from the subchannels. For example, if the gain of subchannel $n$ for user $m$ is $g_m^n$ and assigned power level of this subchannel is $p_m^n$, then the received power from this subchannel is $p_m^ng_m^n$.  We use the notation ${{\bm{\Omega}}'}_m {\succ} {{\bm{\Omega}}''}_m$ to imply that user $m$ wants to have the subchannels in subset ${{\bm{\Omega}}'}_m$ than the subchannels in subset ${{\bm{\Omega}}''}_m$, where ${{\bm{\Omega}}'}_m \subseteq \textbf{N}$ and ${{\bm{\Omega}}''}_m \subseteq \textbf{N}$. Similar analogy can be made for any subchannel $n$ in set \textbf{N}. Preference of each subchannel $n$ is based on the overall benefit (i.e., throughput) of the system. For example, if user $m$ chooses subchannel $n$, this subchannel only accepts this user if and only the system performance is enhanced by this allocation. 

To solve our subchannel-user mapping problem, we are interested to look at a stable solution, in which there are no players that are not matched to one another but they all prefer to be partners. Since subchannel players give preference to the overall throughput of the system while choosing partners from set \textbf{M}, stable solution is envisioned to be the optimal solution for this problem. In many-to-many matching models~\cite{ARoth1984}, many stability concepts can be considered depending on the number of players that can improve their utility by forming new partners among one another. However, due to the large number of players ($\textbf{M} \cup \textbf{N}$) in our problem, identifying optimal subset of partners for a player is intractable. Consequently, we choose to solve the matching problem by identifying partner one by one from the opposite set. This way of choosing partner in the matching model brings pair-wise stability.

In \textit{Definition 2} and \textit{Definition 3}, we highlight some properties of pair-wise stable matching relation. For the sake of these definitions, we define some notations as follows. Faced with a set $\hat{\textbf{N}} \subseteq \textbf{N}$ of possible partners, a player $m \in \textbf{M}$ can determine which subset of set $\hat{\textbf{N}}$, it wishes to match to. We denote this choice set as $\textbf{C}_m(\hat{\textbf{N}})$. 


\noindent
\textbf{Definition 2:} A matching relation $\mu$ is pairwise stable if there does not exist a pair $(m, n)$ with $m \not\in {\mu}(n)$ and $n \not\in {\mu}(m)$ such that $\phi \in \textbf{C}_m({\mu}(m) \cup \{n\})$ and $\varphi \in \textbf{C}_n({\mu}(n) \cup \{m\})$, and at the same time both $\{\phi\} {\succ}_m {\mu}(m)$ and $\{\varphi\} {\succ}_n {\mu}(n)$ are satisfied.

\noindent
\textbf{Definition 3:} Let $\hat{\textbf{M}}$ is the subset of users in set \textbf{M}.  The preference of subchannel $n$ is called substitutable if there exist users such that $m, m' \in \textbf{C}_n(\hat{\textbf{M}})$, then $m \in \textbf{C}_n(\hat{\textbf{M}} /\ \{m'\})$ is satisfied.

While satisfying the properties of stable many-to-many matching relation, we have proposed an algorithm in \textit{Algorithm~\ref{alg:sc-usr-map}}. Note that in this algorithm, we are interested in pair-wise stability, and hence the players (i.e., users and subchannels) choose their partners one-by-one instead of a subset. We have adopted a few paradigms or strategies in order to bring stability in this matching relation or enhance the overall system performance. The description of the algorithm is as follows. First, ${\bm{\Omega}}_m, m \in \textbf{M}$ are initialized with $\emptyset$, which basically contains the allocated subchannels of user $m \in \textbf{M}$. At the same time, $\textbf{M}_n, n \in \textbf{N}$ are initialized with $\emptyset$ as well. Over the iterations, these sets are filled by the allocated subchannels and users, respectively. At the initialization phase, each user $m \in \textbf{M}$ also constructs its subchannel preference list based on the descending order of their received power level. If the gain of subchannel $n$ for user $m$ is $g_m^n$ and the assigned power level is $p_m^n$, the received power level of this subchannel for this user is $p_m^ng_m^n$. Since we have an assumption that the maximal power level of user $m$ is subdivided equally among its allocated subchannels, the preference list of user $m$ is constructed based on the assumption that $\frac{p_m^{\mbox{max}}}{|{\bm{\Omega}}_m|+1}$ amount of power is reserved for subchannel $n \in \{\textbf{N} /\ {\bm{\Omega}}_m\}$. Then, inside the outer-most loop (between step $3$ and step $34$), if no assignment is possible inside the second outer loop (between line $4$ and line $33$), the algorithm terminates\footnote{At this point, it is assumed that the system has reached a stable situation or the performance improvement is no longer possible.}. Inside the inner-most loop (between step $5$ and step $32$), each user $m$ chooses its most preferred unallocated subchannel $n$. At this point, two conditions are possible. The first condition is that the number of allocated users on subchannel $n$ can be less than $K$ (maximum allowable number of users per subchannel), and the second condition is the other case. If the first condition is true, we can apply two strategies for this subchannel-user assignment: either user $m$ is substituted by one of the existing users (e.g., $m' \in \textbf{M}_n$) on subchannel $n$, or user $m$ can be added to this subchannel. Each of these strategies is inserted to strategy set \textbf{S} (which was initialized before initiating the loop). Whereas, for the second case, only addition strategy is possible. After filling the strategy set \textbf{S} no matter the number of allocated users on subchannel $n$ is less than or equal to $K$, the elements of \textbf{S} are filtered based on some criterion, which is as follows. If strategy $s$ is a substitution policy, let $m'$ is to-be-replaced user, and hence $\textbf{N}' = {\bm{\Omega}}_{m} \cup {\bm{\Omega}}_{m'}$ is the set of affected subchannels. Moreover, let $thr'$ be the total computed throughput (following (\ref{eq:Rn})) of the subchannels in set $\textbf{N}'$ before applying strategy $s$. Then, after applying strategy $s$ and adjusting the power level of user $m$ and $m'$ in set $\textbf{N}'$, in the similar manner, the total throughput is computed (denoted by $thr$). Strategy $s$ is only added to set \textbf{CS} if and only if this throughput (due to applying this strategy) is bigger than $thr'$. For the addition strategy, the affected subchannels are the ones in set ${\bm{\Omega}}_m$, and hence $\textbf{N}' = {\bm{\Omega}}_m$. For this strategy, in the similar manner, set \textbf{S} is filtered and set \textbf{CS} is updated. Finally, $s_{Best}$ strategy is chosen based on the total throughput each strategy incurs. If $s_{Best}$ is empty, the inner-most loop continues, and the next user is chosen from set \textbf{M} for building its possible strategy set. If $s_{Best}$ is not empty, the corresponding strategy is executed. As a result, set ${\bm{\Omega}}_m$, set ${\bm{\Omega}}_{m'}$ (only for the substitution strategy), and set $\textbf{M}_n$ are updated. The power level of user $m$ for the subchannels in set ${\bm{\Omega}}_m$ are adjusted, and its preference list is updated as well due to the updated power level. For only substitution strategy, the power level of user $m'$ in its affected subchannels and its preference list is updated. By analyzing the algorithm, we conclude \textit{Proposition 1}, \textit{Lemma 1} and \textit{Theorem 1}.

\noindent
\textbf{Proposition 1:} Rejected users by the subchannels are not final. For example, if a user $m \in \textbf{M}$ is rejected by subchannel $n \in \textbf{N}$ at some iteration $i$, at the $i'$th ($i' \ge i$) iteration, it is possible that both $m \in {\mu}(n)$ and $n \in {\mu}(m)$ will appear true.

\begin{algorithm}[H]
\caption{The uplink subchannel-user mapping algorithm using many-to-many matching model.}
\label{alg:sc-usr-map}
\begin{algorithmic}[1]
\STATE ${\bm{\Omega}}_m \leftarrow \emptyset, \forall{m \in \textbf{M}}$; $\textbf{M}_n \leftarrow \emptyset, n \in \textbf{N}$.
\STATE Each user $m \in \textbf{M}$ produces its preference list.


\REPEAT
\FOR{$i \leftarrow 1~\mbox{to}~N$}
\FOR{$m \in \textbf{M}$}
\STATE $n$ $\leftarrow$ The best unallocated subchannel from the preference list of user $m$. 
\STATE $\textbf{S} \leftarrow \emptyset$, $\textbf{CS} \leftarrow \emptyset$, and $\textbf{Thrput} \leftarrow \emptyset$. 
\IF{The number of assigned users on subchannel $n$ is less than $K$}
\STATE Construct each strategy $s$ that is supposed to replace each user $m' \in \textbf{M}_n$ by user $m$, and insert the corresponding $s$ to set \textbf{S}.
\STATE Construct another strategy $s$ that is supposed to add user $m$ to set $\textbf{M}_n$, and insert the corresponding $s$ to set \textbf{S}.
\ELSIF{The number of assigned users on subchannel $n$ is equal to $K$}
\STATE Construct each strategy $s$ that is supposed to replace each user $m' \in \textbf{M}_n$ by user $m$, and insert the corresponding $s$ to set \textbf{S}.
\ENDIF
\FOR{$s \in \textbf{S}$}
\STATE Determine the affected subchannel list $\textbf{N}'$ for strategy $s$.
\STATE Adjust the power level of user $m' \in \textbf{M}_{n'}, \forall{n' \in \textbf{N}'}$.  
\STATE $thr \leftarrow$ The sum-throughput of all subchannels in set $\textbf{N}'$ due to strategy $s$.
\STATE Insert strategy $s$ to \textbf{CS} and insert $thr$ to set \textbf{Thrput} if $thr$ is larger than the sum-throughput of affected subchannel list $\textbf{N}'$ before applying strategy $s$.
\ENDFOR

\STATE $s_{Best} \leftarrow \argmax_{s \in \textbf{CS}}\textbf{Thrput}(s)$.
\IF{$s_{Best}$ is the replacement strategy}
\STATE ${\bm{\Omega}}_m \leftarrow {\bm{\Omega}}_m \cup \{n\}$, ${\bm{\Omega}}_{m'} \leftarrow {\bm{\Omega}}_{m'} /\ \{n\}$, and $\textbf{M}_n \leftarrow (\textbf{M}_n /\ \{m'\}) \cup \{m\}$. \COMMENT{$m'$ is the to-be-replaced user and $m$ is the new user on subchannel $n$}
\STATE Adjust the power level of the subchannels in ${\Omega}_m$ and ${\Omega}_m'$, and update the preference lists of user $m$ and user $m'$.
\ELSIF{$s_{Best}$ is the addition strategy}
\STATE $m$ is the new user on subchannel $n$. 
\STATE ${\Omega}_m \leftarrow {\Omega}_m \cup \{n\}$, and $\textbf{M}_n \leftarrow \textbf{M}_n \cup \{m\}$. \COMMENT{$m$ is the new user on subchannel $n$}
\STATE Adjust the power level of the subchannels in ${\Omega}_m$, and update the preference list of user $m$.
\ENDIF
\IF{$s_{Best}$ is not empty}\STATE Terminate this loop.
\ENDIF
\ENDFOR
\ENDFOR
\UNTIL{The performance enhancement of the system is not possible}
\end{algorithmic}
\end{algorithm}

\noindent
\textbf{Proof: } Let assume $K = 2$, which implies that maximum allowable number of users per subchannel is $2$. Consider that user $1$ and user $2$ are already matched with subchannel $n$ in some iteration $i$, and $p_2^n > p_1^n$ holds. At iteration $i'$ ($i' > i$), user $3$ has come to obtain subchannel $n$ with the power level $p_3^n$, and $p_2^n > p_3^n > p_1^n$ holds. For the sake of simplicity, we further assume that $g_1^n = g_2^n = g_3^n = 1$. We know that if $p_2^n > p_3^n > p_1^n$ holds, $\mbox{log}_2(1 + p_2^n + p_3^n) > \mbox{log}_2(1 + p_2^n + p_1^n)$ always satisfies. Therefore, at this stage, based on the replacement strategy our algorithm provides, user $3$ is replaced by user $1$. Since user $1$ is unallocated from subchannel $n$, in other subchannels (of set ${\bm{\Omega}}_1$) to which user $1$ is belonged to, the power level of this user will be increased. This is because the algorithm ensures that the maximal power level $p_1^{\mbox{max}}$ is equally subdivided among the allocated subchannels of user $1$. In the similar manner and for the same reason, user $1$ may further be replaced by some other user on its some other assigned subchannel (in set ${\bm{\Omega}}_1$). At this stage, at iteration $i''$ ($i'' > i'$), user $1$ is able to compete (although rejected already at iteration $i'$) for obtaining the $n$th subchannel again with the increased level of power, denoted by ${\tilde{p}}_1^n$ $({\tilde{p}}_1^n > p_3^n)$. Thus, we prove that user $1$ may come again to choose subchannel $n$, and can be replaced by user $3$. This is due to the fact $\mbox{log}_2(1 + p_2^n + {\tilde{p}}_1^n) > \mbox{log}_2(1 + p_2^n + p_3^n)$ because of ${\tilde{p}}_1^n > p_3^n$. By adopting the strategy in this proposition, the algorithm ensures as better performance as possible for the system.

\noindent
\textbf{Lemma 1:} The subchannel-user mapping algorithm (i.e., \textit{Algorithm~\ref{alg:sc-usr-map}}) is guaranteed to converge to a pair-wise stable matching relation.

\noindent
\textbf{Proof:} We prove this lemma by contradiction. Suppose that there exist a user $m$ and a subchannel $n$ with $m \not\in {\mu}(n)$ and $n \not\in {\mu}(m)$ such that  $\phi \in \textbf{C}_m({\mu}(m) \cup \{n\})$, $\varphi \in \textbf{C}_n({\mu}(n) \cup \{m\})$, and at the same time both $\{\phi\} {\succ}_m {\mu}(m)$ and $\{\varphi\} {\succ}_n {\mu}(n)$ are satisfied. Since $\{n\} {\succ}_m {\mu}(m)$ is true, user $m$ must propose subchannel $n$ in some earlier iteration to be paired with. However, at the same time, both $m \not\in {\mu}(n)$ and $n \not\in {\mu}(m)$ are true. Consequently, at the proposal time of user $m$, either subchannel $n$ had some better preference compared with user $m$ and rejected this user, or accepted this user and then made a replacement with some other user in the latter iteration. Therefore, $m \not\in \textbf{C}_n({\mu}(n) \cup \{m\})$ cannot be a false statement, and hence matching relation $\mu$ cannot be unstable.

\noindent
\textbf{Theorem 1:} \textit{Algorithm~\ref{alg:sc-usr-map}} terminates after finite number of iterations.

\begin{figure}[h!]%
    \centering
    \subfloat[M = 40.\label{fig:convg-40-usrs}]{{\includegraphics[scale=0.3]{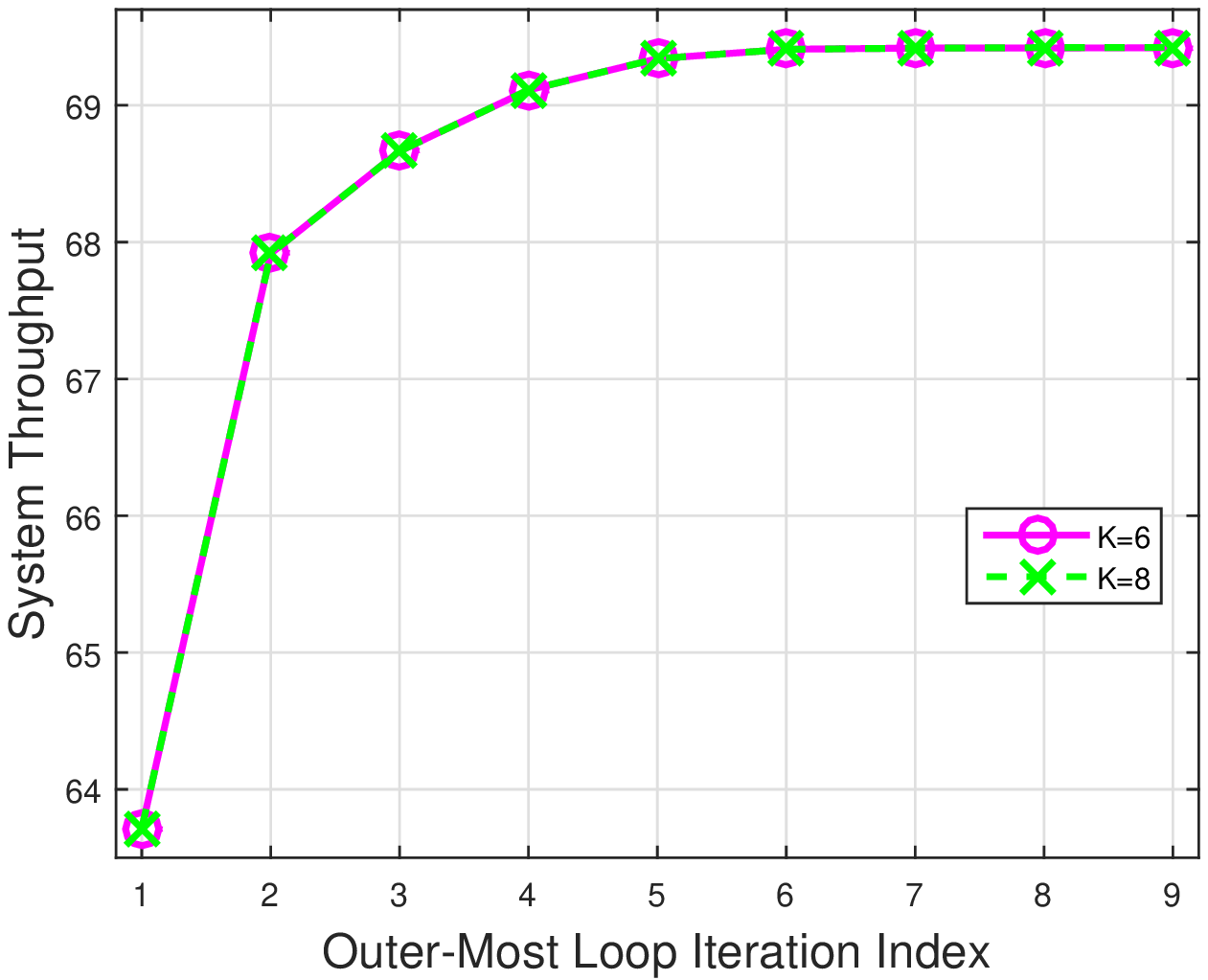} }}%
   ~
    \subfloat[M = 60.\label{fig:convg-40-usrs}]{{\includegraphics[scale=0.3]{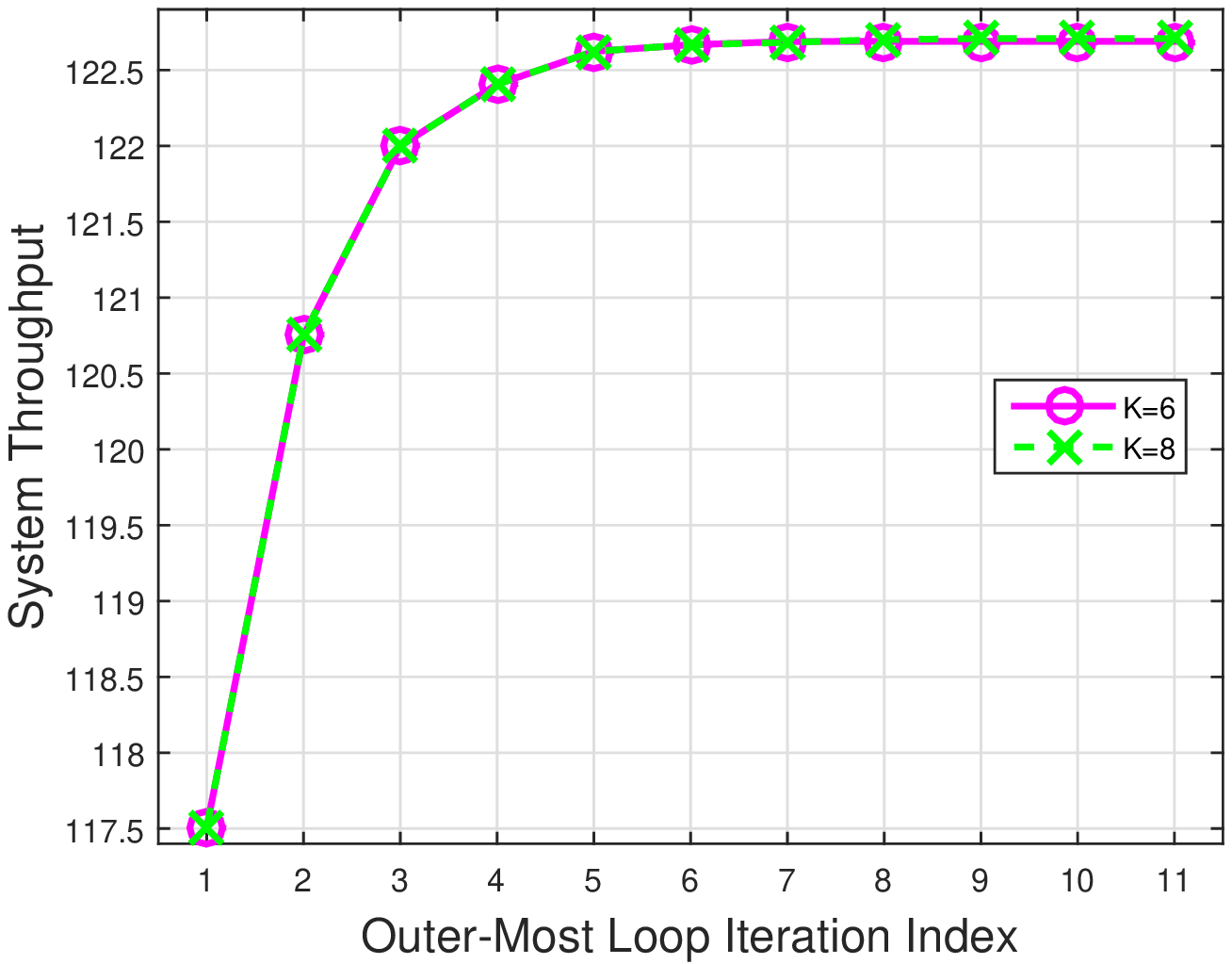} }}%
    \caption{Increasing system throughput as the outer-most loop iteration of \textit{Algorithm~\ref{alg:sc-usr-map}} advances.}%
    \label{fig:convg}%
\end{figure}

\noindent
\textbf{Proof:} \textit{Algorithm~\ref{alg:sc-usr-map}} is proposed to solve the subchannel-user mapping problem, and this is an optimization problem. No matter the problem is convex or non-convex, we know that every optimization problem has a unique global solution. If the problem is non-convex, there might be some local optimal solutions. However, for the convex problem, the solution is unique, which can be assumed as both local and global. As we verify before that the problem discussed herein is non-convex and NP-hard, the global optimal solution requires to search all possible feasible solution spaces, which is computationally intensive and is not feasible to implement in a wireless system, where channel turnaround time is in millisecond/microsecond level. Consequently, we proposed \textit{Algorithm~\ref{alg:sc-usr-map}} to solve this problem. We do not claim that the algorithm always finds the global optimal solution as this is based on the many-to-many matching scheme. In the many-to-many matching model, since the number of players (i.e., users and subchannels) on the both sides is large, in terms of their associations, many combinations are possible as mentioned above. Therefore, we mostly focused on pair-wise stability, and \textit{Algorithm~\ref{alg:sc-usr-map}} is developed based on this concept. The outer-most loop continues if and only if at least one allocation (subchannel-user mapping) is executed. Interestingly, in our algorithm, every allocation enhances the system throughput compared to the throughput before that particular allocation. Step $18$ of the algorithm is the evidence of this statement. Through addition and substitution operations, for a user, we define possible strategies for the tagged user and its preferred subchannel. This implies that the tagged user will be added to its most preferred subchannel or will replace an existing user of that subchannel if only if the system capacity is enhanced by this allocation. Therefore, in every allocation, if the system capacity is enhanced, eventually, the process approaches the convergence state as the solution of the problem exists in the finite domain. Moreover, no matter the problem is convex or non-convex, the global or local optimum point has convex nature. Consequently, after finite number of iterations, the algorithm converges and terminates. In order to provide practical evidence of the convergence event, we have plot Fig.~\ref{fig:convg}. In each subfigure of this figure, we show the system throughput with the increasing outer-most loop iterations for different values of $K$. As observed, in each iteration, the system throughput is increased little by little before reaching the convergence state.

\begin{figure}[h!]%
    \centering
    \subfloat[Results while varying M.\label{fig:its-vs-M}]{{\includegraphics[scale=0.3]{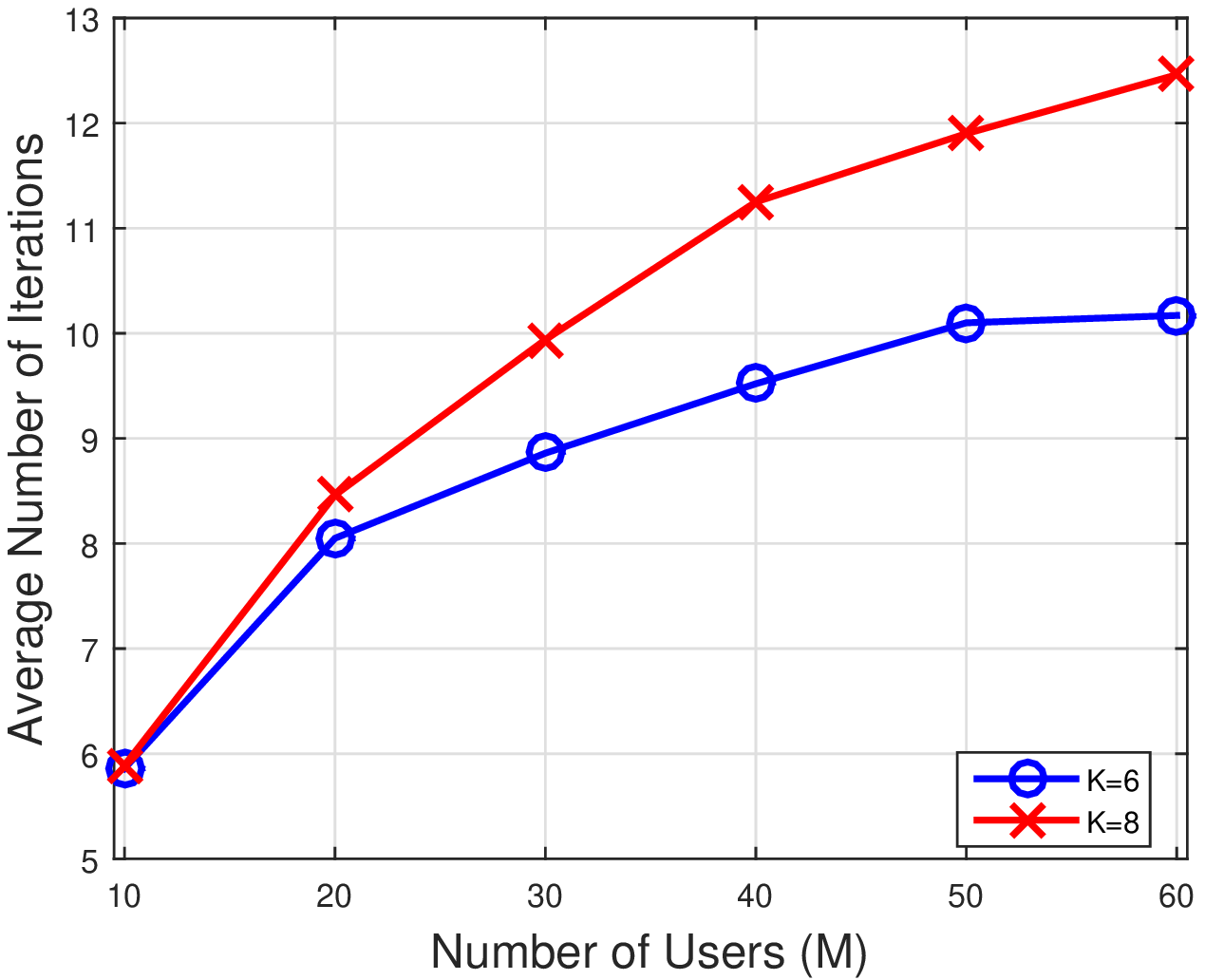} }}%
   ~
    \subfloat[Results while varying K.\label{fig:its-vs-K}]{{\includegraphics[scale=0.3]{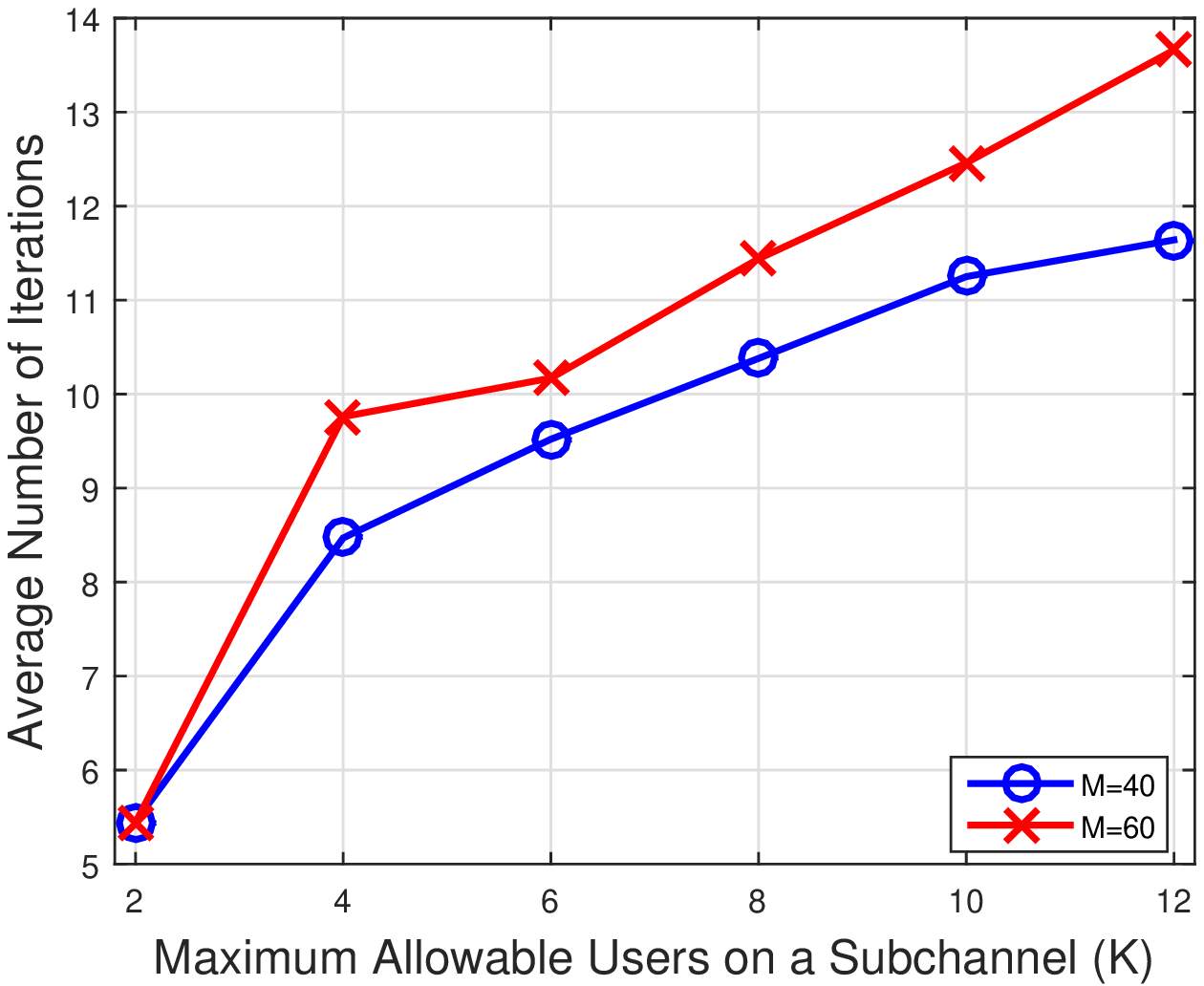} }}%
    \caption{The number of outer-most loop iterations required before achieving stability.}%
    \label{fig:example}%
\end{figure}

In Fig.~\ref{fig:its-vs-M} and Fig.~\ref{fig:its-vs-K}, we present the average number of iterations the outer-most loop runs before reaching the optimal point with the increasing number of users and the increasing value of $K$, respectively. The total number of subchannels in both figures is constant and fixed. Intuitively, given the number of users and the number of subchannels to be matched, the outer-most loop should run the times surrounding the value of $K$. However, when the number of users is less, not necessarily, each subchannel obtains exactly $K$ number of users. Moreover, when the number of users in the system is less, the number of strategies with substitution operation is relatively less compared to the case when the number of users in the system is higher. In this case, the outer-most loop terminates in less number of iterations, which is obvious in Fig.~\ref{fig:its-vs-M}. On the other hand, when the value of $K$ is lower, for the given number of users and subchannels in the system, each subchannel obtains relatively less number of users compared to the case with larger value of $K$. Therefore, due to the restriction on the less number of users allocated to each subchannel, the number of strategies accompanied with substitution and addition operations is less in this case as well similar to the other case. Consequently, the outer-most loop requires less number of iterations to run before reaching the optimal possible point of the system. Fig.~\ref{fig:its-vs-K} presents the corresponding observation in this context.


\noindent
\textbf{Computational Complexity of \textit{Algorithm~\ref{alg:sc-usr-map}}:} The joint worst case complexity of the inner two loops of the algorithm is $\mbox{O}(MN)$. Inside these loops, all operations occur in constant time, and so we can ignore the complexity of these operations. Mainly, the running time of the outer-most loop dominates the computation time of the entire algorithm. As shown in Fig.~\ref{fig:its-vs-M} and Fig.~\ref{fig:its-vs-K}, the iterations of this outer-most loop is proportional to the number of users ($M$) and the maximum allowable number of user per subchannel ($K$), but this is not a very large number. Therefore, we conclude that the algorithm has polynomial time complexity. Whereas, the complexity of the brute-force searching operation to obtain the optimal solution is $\mbox{O}(\frac{M^N {\times}(M-1)^N {\times} \cdots}{1^N {\times} 2^N {\times}\cdots}) \approx \mbox{O}(M^N)$, which is an order of exponential series.

\subsection{Power Allocation Schemes}   
From \textit{Algorithm~\ref{alg:sc-usr-map}}, we know the subchannel-user mapping information, i.e., $\textbf{M}_n, n \in \textbf{N}$ and ${\bm{\Omega}}_m, m \in \textbf{M}$. This information is derived based on the assumption that the maximal power level $p_m^{\mbox{max}}$ of user $m$ is equally subdivided among its allocated subchannels, i.e., $p_m^n = p_m^{\mbox{max}}/|{\bm{\Omega}}_m|, m \in \textbf{M}$. However, in (\ref{eq:opt-prob1}), we see that the instantaneous rate of user $m$ on subchannel $n$ is positively proportional to $p_m^n$ and inversely proportion to the interference power level of other users, i.e., $p_i^n, i \in \textbf{M}_n^m$. Consequently, even if the information about $\textbf{M}_n, n \in \textbf{N}$ and ${\bm{\Omega}}_m, m \in \textbf{M}$ is known, the power allocation across all subchannel-user slots, i.e., $p_m^n, m \in \textbf{M}, n \in \textbf{N}$, is an optimization problem. Consequently, the next objective of this resource allocation scheme is to allocate power across all subchannel-user slots optimally. We have adopted two approaches to solve this power allocation problem, the individual description of which is provided in the following discussions.


\subsubsection{Iterative Water-Filling Algorithm}

Given that subchannel-user mapping information, $\textbf{M}_n, n \in \textbf{N}$ and $\bm{\Omega}_m, m \in \textbf{M}$ are known, the power allocation problem can be written as

\begin{eqnarray}
\label{eq:it-water-fill}
\nonumber & \displaystyle\max_{\{p_m^n\}_{m \in \textbf{M}_n, n \in \textbf{N}}}\displaystyle\sum_{n \in \textbf{N}}\mbox{log}_2\left(1 + \displaystyle\sum_{m \in \textbf{M}_n}p_m^ng_m^n\right), \\
& \mbox{s.t.,}~\displaystyle\sum_{n \in {\bm{\Omega}}_m}p_m^n \le p_m^{\mbox{max}},~\forall{m \in \textbf{M}},
\end{eqnarray}

\noindent
where the objective function is the simplified version of the objective function in~(\ref{eq:opt-prob1}). This is actually a multi-user water-filling problem. Using the dual decomposition method~\cite{DPBertsekas1999}, the solution of this problem is described as follows. Taking the dual variables ${\lambda}_m, m \in \textbf{M}$, the Lagrangian of the problem in~(\ref{eq:it-water-fill}) can be written as

\begin{eqnarray}
\label{eq:it-water-fill-dual}
\nonumber &L(\{p_m^n\}_{m \in \textbf{M}_n, n \in \textbf{N}}) = \displaystyle\sum_{n \in \textbf{N}}\mbox{log}_2\left(1 + \displaystyle\sum_{m \in \textbf{M}_n}p_m^ng_m^n\right), \\
&+ \displaystyle\sum_{m \in \textbf{M}}{\lambda}_m\left(p_m^{\mbox{max}} - \displaystyle\sum_{n \in {\Omega}_m}p_m^n\right).
\end{eqnarray}

Taking the derivative of the Lagrangian in~(\ref{eq:it-water-fill-dual}) with respect to $p_m^n, m \in \textbf{M}_n, n \in \textbf{N}$, we obtain

\begin{equation}
\label{eq:it-water-fill-sol}
p_m^n = \frac{1}{{\lambda}_m} - \frac{1}{g_m^n}\left(1 + \displaystyle\sum_{i \in \textbf{M}_n, i \neq m}p_i^ng_i^n\right),~\forall{m \in \textbf{M}_n}, \forall{n \in \textbf{N}}.
\end{equation}

If we compare the solution in~(\ref{eq:it-water-fill-sol}) with the single-user water-filling solution $p_m^n = 1/{{\lambda}_m} - 1/{g_m^n}$, it is obvious that the optimal power level of one user considers the received power of other users as noise. Based on this intuition, we develop an iterative algorithm in \textit{Algorithm~\ref{alg:power-alloc}} to solve this power allocation problem. The algorithm works as follows. First, power level of all users over all subchannels are initialized. Then, for each user, water-filling power allocation problem is solved assuming the power level of other users as noise. Once the single-user water-filling solutions are obtained for all users, the resultant solutions of all users are replaced by the previously initialized power levels. This operation is continued until the performance of the system appears to be saturated.

While solving the single-user water-filling problem in \textit{Algorithm~\ref{alg:power-alloc}}, typically, bisection search is applied to obtain the optimal value of ${\lambda}_m, m \in \textbf{M}$. If the proper interval of the bisection search is not chosen, running time of the bisection search is huge. Moreover, the accuracy of the solution obtained from the bisection search is greatly dependent on the precision level of ${\lambda}_m$ as this is a variable with continuous nature. Therefore, to obtain the optimal value of ${\lambda}_m$, we have developed a low complexity procedure in \textit{Algorithm~\ref{alg:lambda}}. In this algorithm, for user $m$, $\textbf{A}_m$ is a vector, the elements of which are $\left[\frac{1}{g_m^n}\left(1 + \sum_{i \in \textbf{M}_n, i \neq m}p_i^ng_i^n\right), n \in {\bm{\Omega}}_m\right]$. The insights of this procedure is developed based on the following relation for individual user $m$

\begin{eqnarray}
&\displaystyle\sum_{n \in {\bm{\Omega}}_m}\frac{1}{{\lambda}_m} - \frac{1}{g_m^n}\left(1 + \displaystyle\sum_{i \in \textbf{M}_n, i \neq m}p_i^ng_i^n\right) = p_m^{\mbox{max}} \\
&{\lambda}_m = \frac{|{\bm{\Omega}}_m|}{p_m^{\mbox{max}} + \frac{1}{g_m^n}\left(1 + \displaystyle\sum_{i \in \textbf{M}_n, i \neq m}p_i^ng_i^n\right)}.
\end{eqnarray}

\begin{algorithm}[h!]
\caption{The iterative water-filling algorithm to calculate optimal $\{p_m^n\}_{m \in \textbf{M}_m, n \in \textbf{N}}$.}
\label{alg:power-alloc}
\begin{algorithmic}[1]
\STATE $p_m^n \leftarrow 0,~\forall{m \in \textbf{M}_n}, \forall{n \in \textbf{N}}.$
\REPEAT
\FOR{$m \in \textbf{M}$}
\STATE $\displaystyle\argmax_{\{p_m^n\}_{n \in {\bm{\Omega}}_m}}\displaystyle\sum_{n \in {\bm{\Omega}}_m}\mbox{log}_2\left(1 + p_m^ng_m^n + \displaystyle\sum_{i \in \textbf{M}_n, i \neq m}p_i^ng_i^n\right),$
\STATE $\mbox{s.t.,}~\displaystyle\sum_{n \in {\bm{\Omega}}_m}p_m^n \le p_m^{\mbox{max}}$.
\ENDFOR
\STATE Update the values of previous $\{p_m^n\}_{m \in \textbf{M}_n, n \in \textbf{N}}$ by optimal $\{p_m^n\}_{m \in \textbf{M}_n, n \in \textbf{N}}$ obtained from this iteration.
\UNTIL{The performance improvement is not possible}

\end{algorithmic}
\end{algorithm}

\begin{algorithm}[h!]
\caption{The iterative process to calculate optimal $\lambda_m$ for user $m$ in \textit{Algorithm~\ref{alg:power-alloc}}.}
\label{alg:lambda}
\begin{algorithmic}[1]
\STATE The elements of $\textbf{A}_m$ is sorted in the ascending order.
\STATE $j \leftarrow 1, G_{n} \leftarrow 0, G_d \leftarrow 0$.
\REPEAT
     \STATE $G_n \leftarrow G_n + 1$.
     \STATE $G_d \leftarrow G_d + A_m(j)$.
     \STATE $\lambda_m(j) \leftarrow \frac{G_n}{p_m^{\mbox{max}} + G_d}$.
     \IF{$\lambda_m(j) \ge 1/A_m(j+1)$}
        \STATE $\lambda_m^* \leftarrow \lambda_m(j)$.
        \STATE Break the loop.
     \ENDIF
     \STATE $j \leftarrow j + 1$.
\UNTIL{$j \le |{\bm{\Omega}}_m|$}

\end{algorithmic}
\end{algorithm}

\subsubsection{Geometric Programming}

Another way to write the problem in~(\ref{eq:it-water-fill}) is as follows

\begin{eqnarray}
\label{eq:gp-form}
\nonumber & \displaystyle\min_{\{p_m^n\}_{m \in \textbf{M}_n, n \in \textbf{N}}}\mbox{log}_2\frac{1}{\displaystyle\prod_{n \in \textbf{N}}(1 + \displaystyle\sum_{m \in \textbf{M}_n}p_m^ng_m^n)} \\
\nonumber & \approx \displaystyle\min_{\{p_m^n\}_{m \in \textbf{M}_n, n \in \textbf{N}}}\frac{1}{\displaystyle\prod_{n \in \textbf{N}}(1 + \displaystyle\sum_{m \in \textbf{M}_n}p_m^ng_m^n)}, \\
&\mbox{s.t.,}~\displaystyle\sum_{n \in {\bm{\Omega}}_m}p_m^n \le p_m^{\mbox{max}},~\forall{m \in \textbf{M}}.
\end{eqnarray}

GP~\cite{Boyd2007, MChiang2005} is an optimization technique that can solve some non-convex problem by adopting some transformation on the optimization variables. The objective and constraint functions with which GP deals are posynomials and monomials. The objective function in~(\ref{eq:gp-form}) is the ratio of two posynomials. The ratio of two posynomials is not a posynomial\footnote{The ratio of a posynomial and a monomial is a posynomial.}, and hence this problem is still not amenable to GP. However, there are some heuristics, such as single condensation method, double condensation method~\cite{ MChiang2005} that can be used to make the problem amenable to GP. We have adopted single condensation method to solve this problem. According to this method, the denominator (which is a posynomial) has to be approximated by some monomial. If we denote the denominator of the optimization problem by $G(\{p_m^n\}_{m \in \textbf{M}_n, n \in \textbf{N}})$, the approximated monomial of this function is

\begin{eqnarray}
\nonumber & G(\{p_m^n\}_{m \in \textbf{M}_n, n \in \textbf{N}}) = \displaystyle\prod_{n \in \textbf{N}}(1 + \displaystyle\sum_{m \in \textbf{M}_n}p_m^ng_m^n) \\
& \approx {\lambda}\displaystyle\prod_{m \in \textbf{M}_n, n \in \textbf{N}}(p_m^n)^{a_m^n}.
\end{eqnarray}

\noindent
where $a_m^n, m \in \textbf{M}_n, n \in \textbf{N}$ and $\lambda$ are auxiliary variables. Given the values of $\{p_m^n\}_{m \in \textbf{M}_n, n \in \textbf{N}}$,  the values of the auxiliary variables can be obtained as follows.

\begin{equation}
a_m^n = \frac{p_m^n}{G(\{p_m^n\}_{m \in \textbf{M}_n, n \in \textbf{N}})}\frac{{\partial}G(\{p_m^n\}_{m \in \textbf{M}_n, n \in \textbf{N}})}{{\partial}p_m^n},
\end{equation}

\begin{equation}
\lambda = \frac{G(\{p_m^n\}_{m \in \textbf{M}_n, n \in \textbf{N}})}{\displaystyle\prod_{m \in \textbf{M}_n, n \in \textbf{N}}(p_m^n)^{a_m^n}},
\end{equation}

\noindent
and

\begin{equation}
\frac{{\partial}G(\{p_m^n\}_{m \in \textbf{M}_n, n \in \textbf{N}})}{{\partial}p_m^n} = g_m^n\displaystyle\prod_{n' \in \textbf{N}, n' \neq n}(1 + \displaystyle\sum_{m' \in \textbf{M}_{n'}}p_{m'}^{n'}g_{m'}^{n'}).
\end{equation}

Given some initial values of $\{p_m^n\}_{m \in \textbf{M}_n, n \in \textbf{N}}$, we require an iterative process in order to obtain the optimal values of these variables step by step. The steps of this iterative process are provided in Section III of~\cite{RRuby20152}. The final values of the variables $p_m^n, m \in \textbf{M}_n, n \in \textbf{N}$, obtained in the last iteration of the iterative process, is the solution of our defined optimization problem.

 
\section{Performance Evaluation}
\label{sec:eval}

In this section, we evaluate the performance of our proposed uplink resource allocation scheme via extensive simulation. The detailed system setup and simulation settings are provided in the following subsection. Then, we exhibit the results obtained from the conducted simulation to verify the effectiveness of our scheme.   

\subsection{Simulation Setup}

The cellular network, that we consider, has a circle-like shape. Since our proposed resource allocation scheme did not capture the interference from neighboring cells, we assume that the cellular network is isolated. The base station is located at the center of the cell, and the users are uniformly distributed in a circular range with $500$ m radius. We set the minimum distance between users to $40$ m, and the minimum distance from the users to the base station to $50$ m. As mentioned previously, time is divided into frames. Each time frame is equivalent to $1$ s, and during this frame, spectrum is subdivided among $20$ subchannels and these are available to be allocated among $M$ users.

Each subchannel is assumed to have $200$ KHz bandwidth. According to~\cite{XQiu1999}, the theoretical limit of the channel capacity is given by $\frac{-1.5}{\mbox{ln}(5P_b)}$, where $P_b$ denotes the Bit Error Rate (BER). BER for the channel is configured as $10^{−6}$. The channel between the base station and a user is affected by shadow and Rayleigh fading. Shadowing effect follows log-normal distribution with variance $3.76$. In order to calculate log-normal shadowing effect of a subchannel, we assume the reference distance as $1$ km and the SNR for this reference distance is $28$ dB.  Reference shadowing effect has also the log normal
distribution with variance $3.76$. Rayleigh fading effect for each user over a subchannel follows Rayleigh distribution with zero mean and $10$ scale factor.  Using all these parameters, the gain of each subchannel for a user towards the base station is computed following ($22$) in~\cite{RRuby2015}. The maximal power level of all users is set to $30$ W. SC-coded signal on each subchannel at the base station is decoded following the SIC technique in~\cite{SVanka2012, NIMiridakis2013}.

In addition to implement our proposed resource allocation scheme, we have implemented relevant other algorithms~\cite{MMollanoori2014, MAlImari2015} proposed in the literature already. For example, in~\cite{MMollanoori2014}, the authors proposed two heuristics in order to maximize the overall capacity and proportional fairness across the system. In the figures demonstrated in the following subsection, these are referred by Alg. 1~\cite{MMollanoori2014} and Alg. 2~\cite{MMollanoori2014}. The algorithm proposed in~\cite{MAlImari2015} is referred by Alg.~\cite{MAlImari2015}. Since we have adopted two techniques in order to allocate power to all subchannels across all users, while referring our algorithm, we use IWF and GP for iterative water-filling and GP power allocation schemes, respectively. Furthermore, obtaining the optimal subchannel-user mapping information is computationally intensive for a large-scale system, and hence we apply brute-force search for a system with $10$ and $20$ users. In the following subsection, for each data point, we conduct the simulation over $10000$ time frames.

\subsection{Simulation Results}

\begin{figure}
  \begin{center}
    \includegraphics[width=0.8\columnwidth]{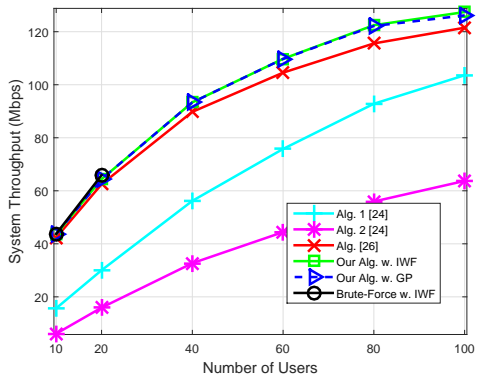}
    \caption{Comparison of system throughput with the increasing number of users.}
    \label{fig:systhr-vs-M}
  \end{center}
\end{figure}

\begin{figure}
  \begin{center}
    \includegraphics[width=0.8\columnwidth]{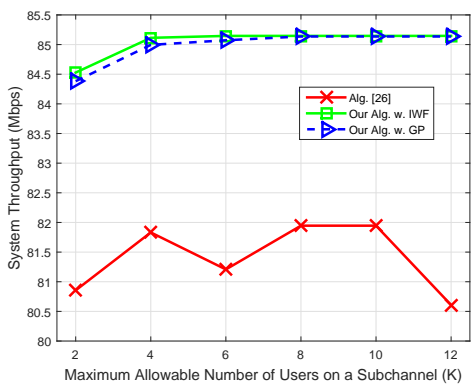}
    \caption{Comparison of system throughput with the increasing value of K.}
    \label{fig:systhr-vs-K}
  \end{center}
\end{figure}

In Fig.~\ref{fig:systhr-vs-M}, we show the increasing system capacity with the increasing number of users. This is a natural trend. The more the number of users in the system the more the overall capacity. In this figure, we set $K$ to $6$. No matter the number of users in the system, our proposed scheme always outperforms existing other algorithms. Actually the algorithms proposed in~\cite{MMollanoori2014} have an assumption that each user can get only one subchannel, which is the main reason of such degraded performance compared to other algorithms. If one user obtains only one subchannel, that assumption fails to exploit the multi-user-channel diversity of wireless systems, and consequently the overall capacity of the system is much lower. On the other hand, the total capacity obtained by the algorithm that is designed to maximize the overall capacity should be larger than that obtained by the algorithm which is designed to maximize the proportional fairness. Fairness of the system is always achieved by sacrificing the capacity of the system.

The algorithm proposed in~\cite{MAlImari2015} has very close performance compared to ours. This algorithm is designed based on iterative water-filling algorithm. The idea of the algorithm is as follows. First, it is assumed that all users are allocated to all subchannels. On this setup, the iterative water-filling algorithm is applied for the power allocation. Obviously, the subchannel-user slot which has the worst gain, obtains the least amount of power. Consequently, the corresponding user is unallocated from the corresponding subchannel. Then, again iterative water-filling algorithm is applied until the constraint that, each subchannel obtains exactly $K$ number of users, is met. At this point, this algorithm is terminated. From the nature and design, it is obvious that the algorithm only keeps the users to a subchannel which have relatively better gain compared to other users. Better gain of a subchannel for a user implies, that user obtains more power on that subchannel. However, if a subchannel has a number of allocated users with relatively better level of power, it does not necessarily enhance the capacity of that subchannel. This is because SIC technique considers the power level of other users as interference level while decoding the signal of one user. As a result, such mechanism of the algorithm does not enhance the system capacity. There must be some other algorithm that should solve the subchannel-user mapping problem in such a way that the drawback of this algorithm can be overcome.

Consequently, we have proposed an efficient algorithm to solve the subchannel-user mapping problem based on the many-to-many matching model. The algorithm is designed in such a way that in every iteration, the system capacity is enhanced little by little. The algorithm terminates only when the performance of the system cannot be enhanced anymore. Based on the first requirement of many-to-many matching model, each user constructs its preference list in the descending order of the received power achieved from the subchannels. Then, each user only wants to obtain its most preferred subchannel, and the corresponding subchannel either adds this user or substitutes the existing user if and only if the sum-capacity of that subchannel and other affected subchannels is enhanced. Note that while solving the subchannel-user mapping problem, it is assumed that each user subdivides its maximal power equally among its allocated subchannels. Furthermore, even if the value of $K$ is large, exactly $K$ number of users allocated to a subchannel (especially the one with worse channel) not necessarily enhances the system throughput. The algorithm adds additional user to a subchannel if that user improves the sum-throughput of that subchannel and other affected subchannels. All these design mechanisms allow the algorithm to overcome the drawbacks of the algorithm in~\cite{MAlImari2015}, and outperforms it. Although for the power allocation of our scheme, we have adopted two techniques, iterative water-filling algorithm outperforms GP technique. Moreover, the way we have implemented iterative water-filling algorithm such that it has much less computational complexity compared to the other one. Therefore, we recommend iterative water-filling as the power allocation technique for our proposed resource allocation scheme.

For $M = 40$ in the system, Fig.~\ref{fig:systhr-vs-K} depicts the increasing system capacity with the increasing value of $K$. It is apparent that the more the number of users in a subchannel, the more enhanced the system capacity is. However, not necessarily the same $K$ number of users on every subchannel enhances the system capacity. This phenomenon particularly happens for the subchannel which has worse condition due to the subdivision of limited power level of each user among its allocated subchannels. Therefore, our algorithm assigns less number of users to some subchannels whenever necessary. However, the algorithm in~\cite{MAlImari2015} blindly assigns exactly $K$ number of users to each subchannel without giving attention to the performance of the system. This statement has been well-proved in this figure. 

\begin{figure}
  \begin{center}
    \includegraphics[width=0.8\columnwidth]{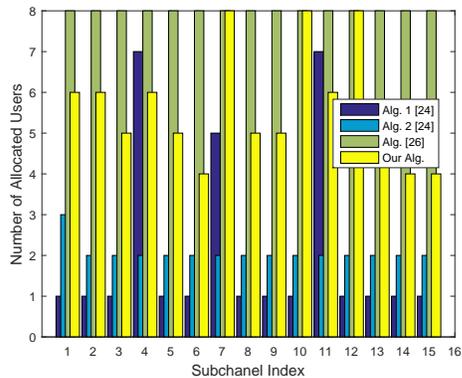}
    \caption{Comparison of allocated users to each individual subchannel when M = 40.}
    \label{fig:N-vs-M}
  \end{center}
\end{figure}

\begin{figure}
  \begin{center}
    \includegraphics[width=0.8\columnwidth]{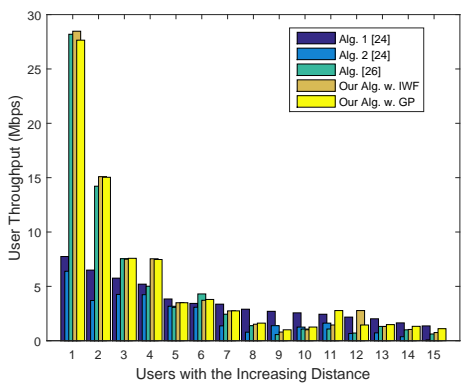}
    \caption{Comparison of throughput for each individual user when M = 40.}
    \label{fig:thr-vs-M}
  \end{center}
\end{figure}

\begin{figure}
  \begin{center}
    \includegraphics[width=0.8\columnwidth]{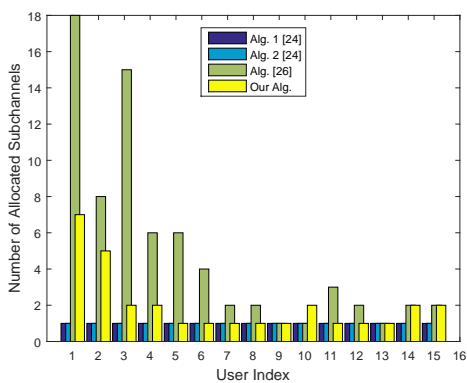}
    \caption{Comparison of allocated subchannels to each individual user when M = 40.}
    \label{fig:scs-vs-M}
  \end{center}
\end{figure}

\begin{figure}
  \begin{center}
    \includegraphics[width=0.8\columnwidth]{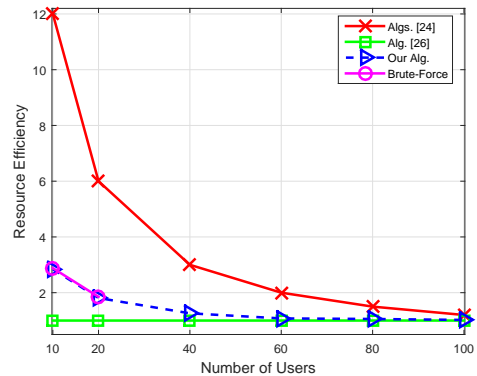}
    \caption{Comparison of resource efficiency with the increasing number of users.}
    \label{fig:rsceff-vs-M}
  \end{center}
\end{figure}

\begin{figure}
  \begin{center}
    \includegraphics[width=0.8\columnwidth]{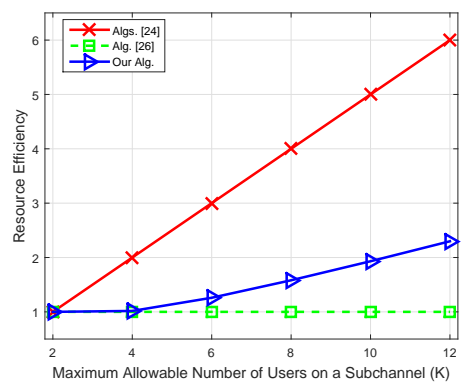}
    \caption{Comparison of resource efficiency with the increasing value of K.}
    \label{fig:rsceff-vs-K}
  \end{center}
\end{figure}

The evidence that our algorithm not necessarily assigns exactly $K$ number of users to all subchannels is strengthened in Fig.~\ref{fig:N-vs-M}. In this figure, $K$ is set to $8$. The subchannel which has relatively better gain can have larger number of users compared to other subchannels. On the other hand, the algorithm in~\cite{MAlImari2015} always assigns $K$ number of users to all subchannels. The algorithms in~\cite{MMollanoori2014} assign the user to a subchannel which has the lowest interference level and each user only obtains one subchannel. Therefore, these two algorithms also assign less number of users compared to the value of $K$.

In Fig.~\ref{fig:thr-vs-M} and Fig.~\ref{fig:scs-vs-M}, we provide some detailed information about our algorithm to justify its superiority. In these figures, $K$ is set to $8$ as well. Since our algorithm is based on the consideration to maximize the system capacity, the user who has the best gain on all subchannels, should have the highest throughput. At the same time, since Rayleigh fading effect is statistically similar for all users over all subchannels, the user closest to the base station should have the highest throughput, and so thus observed in Fig.~\ref{fig:thr-vs-M}. There are some exceptions as well due to the random nature of Rayleigh fading effect, such as users $9, 10, 11$ and $12$. Other than some exceptions, the throughput of the users have the decreasing trend with the incresing distance from the base station. Due to the aforementioned justifications, other algorithms incur less throughput for all users except the one in~\cite{MAlImari2015}. Previously, we claimed that our proposed algorithm is not necessarrily globally optimal, and therefore the algorithm in~\cite{MAlImari2015} incurs larger throughput for one or two users. However, it is obvious in Fig.~\ref{fig:thr-vs-M} that the throughput of more number of users incurred by our algorithm are better compared to the one in~\cite{MAlImari2015}. One of the design insights of our proposed resource allocation scheme is to exploit multi-user-channel diversity of wireless systems. Consequently, by assigning more number of subchannels to each user, it is possible to enhance the system throughput, and so thus our algorithm does (as depicted in Fig.~\ref{fig:scs-vs-M}). However, the algorithms in~\cite{MMollanoori2014} assign only one subchannel to a user, and hence their performance is much worse compared to others.

In Fig.~\ref{fig:rsceff-vs-M} and Fig.~\ref{fig:rsceff-vs-K}, we further justify that our algorithm is superior in terms of resource efficiency as well. In these two figures, we plot the ratio of total subchannel-user slots and the number of used subchannel-user slots with the increasing number of users and with the increasing value of $K$, respectively. If the number of used subchannel-user slots is denoted by $D~(i.e., \sum_{n \in \textbf{N}}|\textbf{M}_n|)$, the metric on the Y-axis of these two figures is $\frac{KN}{D}$. It is obvious that the more the number of users and the larger the value of $K$, the more subchannels are used to support more users and to enhance the system throughput. Previously, we observed that the overall throughput achieved by the algorithm in~\cite{MAlImari2015} has close performance to ours, however using more resource slots in the system as depicted in Fig.~\ref{fig:rsceff-vs-M} and Fig.~\ref{fig:rsceff-vs-K}. On the other hand, although the algorithms in~\cite{MMollanoori2014} have the highest resource efficiency, they have the worst overall performance as proved in the previous results. These algorithms incur the highest performance in this case because of assigning only one subchannel to each user. Whereas, to exploit multi-user-subchannel diversity of wireless systems, our scheme assigns more subchannels to the users, and consequently achieves the highest overall performance.

\section{Conclusion and Future Work}
\label{sec:concl}

In this paper, we proposed an uplink resource allocation scheme of a NOMA system, in which the spectrum is divided into multiple subchannels. The objective of our resource allocation scheme is to allocate power and subchannels across the users of the system. Due to the discrete nature of subchannels and the characteristics of NOMA systems, the problem is NP-hard and non-convex. Since the optimal solution of the problem is intractable, we solved the problem in two steps. First, the subchannel-user mapping problem was solved using many-to-many matching model. Then, iterative water-filling algorithm and GP technique were applied to allocate power optimally across all assigned subchannels and users of the system. We conducted extensive simulation to verify the effectiveness of our proposed resource allocation scheme comparing with other existing works in the literature.

One of our previous experience~\cite{LZhou2014} while dealing with interference is, reducing the allocated power level on the subchannels may bring better performance for the system compared to the case when the maximal power of the users are used. This is what exactly observed while allocating subchannels and power among the users in an OFDMA-based network surrounded by many neighboring cells. In such systems, transmission of users in one cell causes interference for the transmission of users in other cells. Since in NOMA systems, multiple users are superimposed on the same subchannel, power level of one user may cause interference for other users. Both water-filling algorithm and GP technique use full power of a user to its allocated subchannels. However, we believe that using less power, it might be possible to enhance the system performance in terms of both capacity and resource efficiency. Another assumption of our work is, the network for which we proposed the uplink resource allocation scheme, is isolated and does not have any neighboring cell. Such assumption of the network is equivalent to ignoring the interference from neighboring cells. However, interference from the neighboring cells is a crucial factor especially for the uplink case in this context. As of our future work, we would like to continue the research in this direction in order to obtain more promising and useful results.


\begin{thebibliography}{10}
\providecommand{\url}[1]{#1}
\csname url@samestyle\endcsname
\providecommand{\newblock}{\relax}
\providecommand{\bibinfo}[2]{#2}
\providecommand{\BIBentrySTDinterwordspacing}{\spaceskip=0pt\relax}
\providecommand{\BIBentryALTinterwordstretchfactor}{4}
\providecommand{\BIBentryALTinterwordspacing}{\spaceskip=\fontdimen2\font plus
\BIBentryALTinterwordstretchfactor\fontdimen3\font minus
  \fontdimen4\font\relax}
\providecommand{\BIBforeignlanguage}[2]{{%
\expandafter\ifx\csname l@#1\endcsname\relax
\typeout{** WARNING: IEEEtran.bst: No hyphenation pattern has been}%
\typeout{** loaded for the language `#1'. Using the pattern for}%
\typeout{** the default language instead.}%
\else
\language=\csname l@#1\endcsname
\fi
#2}}
\providecommand{\BIBdecl}{\relax}
\BIBdecl

\bibitem{STekinay2002}
S.~Tekinay, ``Next generation wireless networks,'' \emph{Springer US}, 2002.

\bibitem{Borst2005}
S.~C. Borst, A.~Buvaneswari, L.~M. Drabeck, M.~J. Flanagan, J.~M. Graybeal,
  G.~K. Hampel, M.~Haner, W.~M. MacDonald, P.~A. Polakos, G.~Rittenhouse,
  I.~Saniee, A.~Weiss, and P.~A. Whiting, ``Dynamic optimization in future
  cellular networks,'' \emph{Bell Labs Technical Journal}, vol.~10, pp.
  99--119, 2005.

\bibitem{ArunAgarwal2014}
A.~Agarwal and K.~Agarwal, ``The next generation mobile wireless cellular
  networks: {4G} and beyond,'' \emph{American Journal of Electrical and
  Electronic Engineering}, vol.~2, no.~3, pp. 92--97, 2014.

\bibitem{AAli2015}
\BIBentryALTinterwordspacing
A.~Ali, W.~Hamouda, and M.~Uysal, ``Next generation {M2M} cellular networks:
  Challenges and practical considerations,'' \emph{CoRR}, vol. abs/1506.06216,
  2015. [Online]. Available: \url{http://arxiv.org/abs/1506.06216}
\BIBentrySTDinterwordspacing

\bibitem{SGhosh2005}
S.~Ghosh, K.~Basu, and S.~K. Das, ``An architecture for next-generation radio
  access networks,'' \emph{IEEE Netw.}, vol.~19, no.~5, pp. 35--42, Sept 2005.

\bibitem{VSuryaprakash2016}
V.~Suryaprakash and I.~Malanchini, ``Reliability in future radio access
  networks: From linguistic to quantitative definitions,'' in \emph{Proc. IEEE
  IWQoS}, June 2016, pp. 1--2.

\bibitem{TEdler2014}
T.~Edler and S.~Lundberg, ``Energy efficiency enhancements in radio access
  networks,'' \emph{Ericsson Review}, vol.~21, no.~1, 2014.

\bibitem{Higuchi2013}
K.~Higuchi and Y.~Kishiyama, ``Non-orthogonal access with random beamforming
  and intra-beam {SIC} for cellular {MIMO} downlink,'' in \emph{Proc. IEEE VTC
  Fall}, 2013, pp. 1--5.

\bibitem{YEndo2012}
Y.~Endo, Y.~Kishiyama, and K.~Higuchi, ``Uplink non-orthogonal access with
  {MMSE-SIC} in the presence of inter-cell interference,'' in \emph{Proc.
  ISWCS}, Aug 2012, pp. 261--265.

\bibitem{JUmehara2012}
J.~Umehara, Y.~Kishiyama, and K.~Higuchi, ``Enhancing user fairness in
  non-orthogonal access with successive interference cancellation for cellular
  downlink,'' in \emph{Proc. IEEE ICCS}, Nov 2012, pp. 324--328.

\bibitem{NOtao2012}
N.~Otao, Y.~Kishiyama, and K.~Higuchi, ``Performance of non-orthogonal access
  with {SIC} in cellular downlink using proportional fair-based resource
  allocation,'' in \emph{Proc. ISWCS}, Aug 2012, pp. 476--480.

\bibitem{NIMiridakis2013}
N.~I. Miridakis and D.~D. Vergados, ``A survey on the successive interference
  cancellation performance for single-antenna and multiple-antenna {OFDM}
  systems,'' \emph{IEEE Commun. Surveys Tutorials}, vol.~15, no.~1, pp.
  312--335, Jan 2013.

\bibitem{YSaito2013}
Y.~Saito, A.~Benjebbour, Y.~Kishiyama, and T.~Nakamura, ``System-level
  performance evaluation of downlink non-orthogonal multiple access ({NOMA}),''
  in \emph{Proc. IEEE PIMRC}, June 2013, pp. 611--615.

\bibitem{ABenjebbour2013}
A.~Benjebbour, A.~Li, Y.~Saito, Y.~Kishiyama, A.~Harada, and T.~Nakamura,
  ``System-level performance of downlink {NOMA} for future {LTE}
  enhancements,'' in \emph{Proc. IEEE GLOBECOM Workshops}, June 2013, pp.
  611--615.

\bibitem{MRHojeij2015}
M.~R. Hojeij, J.~Farah, C.~A. Nour, and C.~Douillard, ``Resource allocation in
  downlink non-orthogonal multiple access ({NOMA}) for future radio access,''
  in \emph{Proc. IEEE VTC Spring}, May 2015, pp. 1--6.

\bibitem{PParida2014}
P.~Parida and S.~S. Das, ``Power allocation in {OFDM} based {NOMA} systems: A
  {DC} programming approach,'' in \emph{Proc. IEEE Globecom Workshops}, Dec
  2014, pp. 1026--1031.

\bibitem{NVucic2010}
N.~Vucic, S.~Shi, and M.~Schubert, ``{DC} programming approach for resource
  allocation in wireless networks,'' in \emph{Proc. International Symposium on
  Modeling and Optimization in Mobile, Ad Hoc, and Wireless Networks}, May
  2010, pp. 380--386.

\bibitem{SHan2014}
S.~Han, C.~L. I, Z.~Xu, and Q.~Sun, ``Energy efficiency and spectrum efficiency
  co-design: From {NOMA} to network {NOMA},'' \emph{IEEE Commun. Society MMTC
  E-Lett.}, vol.~9, no.~5, pp. 21--24, Sept 2013.

\bibitem{QSun2015}
Q.~Sun, S.~Han, C.~L. I, and Z.~Pan, ``Energy efficiency optimization for
  fading {MIMO} non-orthogonal multiple access systems,'' in \emph{Proc. IEEE
  ICC}, June 2015, pp. 2668--2673.

\bibitem{FFang2016}
F.~Fang, H.~Zhang, J.~Cheng, and V.~C.~M. Leung, ``Energy-efficient resource
  allocation for downlink non-orthogonal multiple access network,'' \emph{IEEE
  Trans. Commun.}, vol.~64, no.~9, pp. 3722--3732, Sept 2016.

\bibitem{JHuang2009}
J.~Huang, V.~G. Subramanian, R.~Agrawal, and R.~Berry, ``Joint scheduling and
  resource allocation in uplink {OFDM} systems for broadband wireless access
  networks,'' \emph{IEEE J. Sel. A. Commun.}, vol.~27, no.~2, pp. 226--234, Feb
  2009.

\bibitem{RRuby2014}
R.~Ruby and V.~Leung, ``Uplink scheduling solution for enhancing throughput and
  fairness in relayed long-term evolution networks,'' \emph{IET Commun.},
  vol.~6, no.~8, pp. 813--825, 2014.

\bibitem{RRuby2015}
R.~Ruby, V.~C.~M. Leung, and D.~G. Michelson, ``Uplink scheduler for
  {SC-FDMA}-based heterogeneous traffic networks with {QoS} assurance and
  guaranteed resource utilization,'' \emph{IEEE Trans. Veh. Technol.}, vol.~64,
  no.~10, pp. 4780--4796, Oct 2015.

\bibitem{MMollanoori2014}
M.~Mollanoori and M.~Ghaderi, ``Uplink scheduling in wireless networks with
  successive interference cancellation,'' \emph{IEEE Trans. Mobile Comput.},
  vol.~13, no.~5, pp. 1132--1144, May 2014.

\bibitem{KKumaran2003}
K.~Kumaran and L.~Qian, ``Scheduling on uplink of {CDMA} packet data network
  with successive interference cancellation,'' in \emph{Proc. IEEE WCNC},
  vol.~3, March 2003, pp. 1645--1650 vol.3.

\bibitem{MAlImari2015}
M.~Al-Imari, P.~Xiao, and M.~A. Imran, ``Receiver and resource allocation
  optimization for uplink {NOMA} in {5G} wireless networks,'' in \emph{Proc.
  ISWCS}, Aug 2015, pp. 151--155.

\bibitem{WeiYu2004}
W.~Yu, W.~Rhee, S.~Boyd, and J.~M. Cioffi, ``Iterative water-filling for
  gaussian vector multiple-access channels,'' \emph{IEEE Trans. Inform.
  Theory}, vol.~50, no.~1, pp. 145--152, Jan 2004.

\bibitem{DPBertsekas1999}
D.~P. Bertsekas, ``Nonlinear programming,'' \emph{Athena Scientific}, 1999.

\bibitem{KHamidouche2014}
K.~Hamidouche, W.~Saad, and M.~Debbah, ``{Many-to-Many} matching games for
  proactive social-caching in wireless small cell networks,'' in \emph{Proc.
  IEEE WiOpt}, May 2014, pp. 569--574.

\bibitem{ARoth1984}
A.~Roth, ``Stability and polarization of interests in job matching,''
  \emph{Econometrica}, vol.~52, no.~1, pp. 47--57, 1984.

\bibitem{MChiang2005}
M.~Chiang, ``Geometric programming for communication systems,'' \emph{Commun.
  Inf. Theory}, vol.~2, no. 1/2, pp. 1--154, Jul. 2005.

\bibitem{DavidTse2005}
D.~Tse and P.~Viswanath, ``Fundamentals of wireless communication,''
  \emph{Cambridge University Press}, 2005.

\bibitem{KMcClaning2000}
K.~McClaning, ``Radio receiver design,'' \emph{Noble Publishing Corporation},
  2000.

\bibitem{HMichiel2001}
M.~Hazewinkel, ``Transmission rate of a channel,'' \emph{Encyclopedia of
  Mathematics, Springer}, 2001.

\bibitem{Boyd2007}
S.~Boyd, S.-J. Kim, L.~Vandenberghe, and A.~Hassibi, ``A tutorial on geometric
  programming,'' \emph{Optimization and Engineering}, vol.~8, no.~1, p.~67,
  2007.

\bibitem{RRuby20152}
R.~Ruby, V.~C.~M. Leung, and D.~G. Michelson, ``Centralized and game
  theoretical solutions of joint source and relay power allocation for {AF}
  relay based network,'' \emph{IEEE Trans. Commun.}, vol.~63, no.~8, pp.
  2848--2863, Aug 2015.

\bibitem{XQiu1999}
X.~Qiu and K.~Chawla, ``On the performance of adaptive modulation in cellular
  systems,'' \emph{IEEE Trans. Commun.}, vol.~47, no.~6, pp. 884--895, Jun
  1999.

\bibitem{SVanka2012}
S.~Vanka, S.~Srinivasa, Z.~Gong, P.~Vizi, K.~Stamatiou, and M.~Haenggi,
  ``Superposition coding strategies: Design and experimental evaluation,''
  \emph{IEEE Trans. Wirel. Commun.}, vol.~11, no.~7, pp. 2628--2639, July 2012.

\bibitem{LZhou2014}
L.~Zhou, R.~Ruby, H.~Zhao, X.~Ji, J.~Wei, and V.~Leung, ``A graph-based
  resource allocation scheme with interference coordination in small cell
  networks,'' in \emph{Proc. IEEE Globecom Workshops}, 2014, pp. 211--218.

\end{thebibliography}


\end{document}